\documentclass[sigplan,10pt]{acmart}
\acmSubmissionID{<141>}

\usepackage{tikz}

\usepackage{amsmath}
\usepackage{subfigure}
\usepackage{pifont}
\usepackage{tikz}
\usepackage{amsmath, amssymb, amsfonts}
\usepackage{multirow}
\usepackage{graphicx}
\usepackage{caption}
\usepackage{makecell}
\usepackage{pifont}
\usepackage{soul}
\usepackage{pifont}
\usepackage[linesnumbered, ruled, noend]{algorithm2e}

\usepackage{xcolor}
\definecolor{darkblue}{HTML}{213D40}
\definecolor{lightblue}{HTML}{D7F0EF}

\usepackage{filecontents}

\AtBeginDocument{%
  }

\usepackage{subfigure}

\setcopyright{acmlicensed}
\acmYear{2026}\copyrightyear{2026}
\setcopyright{cc}
\setcctype[4.0]{by}
\acmConference[EUROSYS '26]{European Conference on Computer Systems}{April 27--30, 2026}{Edinburgh, Scotland Uk}
\acmBooktitle{European Conference on Computer Systems (EUROSYS '26), April 27--30, 2026, Edinburgh, Scotland Uk}
\acmDOI{10.1145/3767295.3769329}
\acmISBN{979-8-4007-2212-7/26/04}

\begin{document}

\title[\textsc{Flux}]{Federated Fine-Tuning of Sparsely-Activated Large Language Models on Resource-Constrained Devices}






\author{Fahao Chen$^{1,2}$, Jie Wan$^{2}$, Peng Li$^{2}$, Zhou Su$^{2}$, Dongxiao Yu$^{1}$}

\affiliation{%
  \institution{$^{1}$Shandong University\quad $^{2}$Xi’an Jiaotong University}
  \country{}
}

\email{chenfh@ieee.org, wanjie_xjtu@stu.xjtu.edu.cn, pengli@xjtu.edu.cn, zhousu@ieee.org, dxyu@sdu.edu.cn}

\renewcommand{\shortauthors}{Fahao Chen, Jie Wan, Peng Li, Zhou Su, and Dongxiao Yu}

\begin{abstract}
Federated fine-tuning of Mixture-of-Experts (MoE)-based large language models (LLMs) is challenging due to their massive computational requirements and the resource constraints of participants. Existing works attempt to fill this gap through model quantization, computation offloading, or expert pruning. However, they cannot achieve desired performance due to impractical system assumptions and a lack of consideration for MoE-specific characteristics.
In this paper, we propose \textsc{Flux}, a system designed to enable federated fine-tuning of MoE-based LLMs across participants with constrained computing resources (e.g., consumer-grade GPUs), aiming to minimize time-to-accuracy. \textsc{Flux} introduces three key innovations: (1) quantization-based local profiling to estimate expert activation with minimal overhead, (2) adaptive layer-aware expert merging to reduce resource consumption while preserving accuracy, and (3) dynamic expert role assignment using an exploration-exploitation strategy to balance tuning and non-tuning experts.
Extensive experiments on LLaMA-MoE and DeepSeek-MoE with multiple benchmark datasets demonstrate that \textsc{Flux} significantly outperforms existing methods, achieving up to 4.75$\times$ speedup in time-to-accuracy. 
\end{abstract}



\maketitle

\section{Introduction}
Sparsely activated large language models (LLMs), particularly those leveraging the Mixture-of-Experts (MoE) architecture, have recently demonstrated strong capability in various NLP tasks~\cite{lepikhin2021gshard, riquelme2021scaling, fedus2022switch}. An MoE-based LLM consists of multiple specialized expert networks, each tailored to handle specific types of input data. Before deployment, it typically undergoes a fine-tuning process, where it is re-trained on domain-specific data to enhance its adaptability and accuracy. In many real-world scenarios, such data is distributed across multiple participants (e.g., organizations or companies) that are either unwilling or legally restricted from sharing their data. This setting naturally aligns with the principles of federated learning~\cite{mcmahan2017communication}, which enables collaborative model training by exchanging model updates instead of raw data, thereby preserving data privacy. Considering this promise, federated LLM fine-tuning has attracted significant attention, with applications in domains such as retail operations optimization~\cite{RevolutionizingEU} and medical report generation~\cite{che2025llm}. Moreover, several open-source frameworks, including NVIDIA FLARE~\cite{roth2022nvidia}, FedAdapter~\cite{cai2023efficient}, and FlowerTune~\cite{gao2025flowertune}, are actively advancing this direction.

Despite extensive research in federated learning, an efficient solution for federated fine-tuning of MoE-based LLMs remains elusive. The primary challenge stems from the massive model size, which often exceeds the computing capabilities of individual participants. 
Therefore, traditional federated learning techniques, such as~\cite{lai2021oort,li2022pyramidfl,cai2023efficient,xu2024fwdllm,ching2024totoro,khan2024float}, which rely on full-model training and parameter exchange, are not directly applicable. Recent studies~\cite{isaksson2022adaptive,guo2021pfl} have explored federated MoE fine-tuning by focusing on expert layers, demonstrating its effectiveness in certain scenarios. However, these methods primarily target small-scale MoE models and assume that the entire model can fit within local computing devices, limiting their scalability and applicability to LLM deployments.

To adapt the MoE-based LLM to constrained computing resources, several optimization techniques can be employed. Quantization is a widely used method for model compression by reducing the precision of model parameters from high-bit floating point representations (e.g., FP32) to lower-bit formats (e.g., INT8). Previous studies have demonstrated that quantization is highly effective for MoE inference~\cite{frantar2024qmoe, yi2025edgemoe}. However, fine-tuning a quantized MoE model often fails to achieve the desired performance, because of accumulated precision errors during fine-tuning. Another approach is to offload some experts to main memory (RAM) and dynamically transfer them to the GPU when needed~\cite{hwang2024pre,kong2024swapmoe,cao2025moe}. However, frequent data transfers between main memory and the GPU introduce substantial latency, significantly slowing down fine-tuning. Some recent studies have explored a more radical approach by pruning ``unimportant'' experts to reduce the model’s computational and memory footprint~\cite{lu2024not,chowdhuryprovably,mei2024fedmoe}. However, our findings indicate that these seemingly ``unimportant'' experts are still crucial for model convergence. Removing them could slow down fine-tuning or even compromise overall performance.

In this paper, we present \textsc{Flux}, a system designed to enable federated fine-tuning of MoE-based LLMs across participants with constrained computing resources (e.g., consumer-grade GPUs), aiming to minimize time-to-accuracy. The core idea of \textsc{Flux} is to let each participant construct a compact MoE model that closely approximates the full model in both tuning behavior and learned representations. Specifically, experts on each participant are categorized into \textit{tuning experts} and \textit{non-tuning ones}, according to their contributions to fine-tuning. Tuning experts are preserved in their original size and updated during tuning, while non-tuning experts are merged and frozen to reduce memory and computational overhead. Note that expert roles, tuning or non-tuning, could dynamically change during the runtime.
Although this idea is promising, there are several critical system-level challenges that must be addressed to ensure efficiency in practical settings, particularly given non-IID data and heterogeneous computing and communication resources across participants.

\textbf{Profiling expert activation}.
The first challenge \textsc{Flux} faces is efficiently profiling expert activation to evaluate their contributions to fine-tuning. Unlike traditional dense models, where all parameters are trained on the entire dataset, MoE models activate only a subset of experts for each input, meaning each expert is trained on a specific subset of tokens. 
By profiling expert activation, we can determine which data should be used to train tuning experts and assess the importance of non-tuning experts.
Some existing works~\cite{zhang2024personalized,mei2024fedmoe} have explored selecting a subset of highly active experts for training. However, they assume that profiling information is readily available. In federated settings, profiling on resource-constrained participants is highly challenging. Running the full MoE model to measure expert activation is computationally infeasible, and delegating this process to third parties compromises data privacy.

\textsc{Flux} addresses this challenge by introducing a \textit{quantization-based local profiling} method, leveraging a quantized MoE model to efficiently estimate expert activation frequency and determine the data processed by each expert. This approach significantly reduces computational overhead while ensuring a reliable assessment of expert relevance for fine-tuning.

\textbf{Merging non-tuning experts}. Non-tuning experts must be efficiently merged to fit within resource constraints while preserving model accuracy. Existing studies~\cite{he2023merging,li2024merge} have demonstrated the benefits of expert merging, but they primarily focus on global model compression, merging experts based solely on activation frequency. However, this simplistic approach overlooks critical factors influencing fine-tuning effectiveness, leaving room for substantial improvement in minimizing its negative impact.

\textsc{Flux} introduces an \textit{adaptive layer-aware merging strategy}, where memory allocation for non-tuning experts is determined based on expert activation distributions and error accumulation across layers. This ensures that layers prone to higher error propagation retain greater expert diversity, mitigating performance degradation. Additionally, instead of relying solely on activation frequency, \textsc{Flux} incorporates token attention scores to assess expert importance during merging, ensuring that the process preserves critical model behaviors.

\textbf{Expert role assignment}.
The final challenge lies in expert role assignment, i.e., determining whether an expert should be tuning or non-tuning, under strict computational and memory constraints at participants. Increasing the number of tuning experts accelerates convergence but prolongs local training time. Moreover, they encroach on the memory space left for non-tuning experts, potentially leading to larger tuning errors. While traditional federated learning has studied a similar problem about participant selection, these solutions cannot be directly applied here, as they do not account for the sparse activation nature of MoE experts.

To address this, \textsc{Flux} defines \textit{expert utility} based on gradient magnitudes and data utilization, ensuring that expert selection prioritizes the most impactful updates. With the objective of maximizing total utility, \textsc{Flux} employs an \textit{exploration-exploitation strategy}, where high-utility experts are prioritized as tuning experts, while a fraction of non-tuning experts is periodically sampled for exploration to refine utility estimates. This adaptive approach balances optimization and adaptability, ensuring that role assignment remains efficient, responsive, and effective over successive fine-tuning rounds.

\noindent\textbf{Results. }We construct a testbed and evaluate \textsc{Flux} on two MoE models, LLaMA-MoE~\cite{llama-moe} and DeepSeek-MoE~\cite{dai2024deepseekmoe}, using four commonly used datasets: Dolly~\cite{DatabricksBlog2023DollyV2}, GSM8K~\cite{cobbe2021gsm8k}, MMLU~\cite{hendryckstest2021}, and PIQA~\cite{Bisk2020}. Extensive experiments demonstrate that \textsc{Flux} effectively accelerates the federated MoE fine-tuning process, achieving a 4.75$\times$ speedup in time-to-accuracy compared to state-of-the-art baselines.

\section{Preliminary and Motivation}
\subsection{Preliminary}
\noindent\textbf{MoE. }The Mixture-of-Experts (MoE) architecture scales the Transformer~\cite{vaswani2017attention} model by introducing MoE layers, each of which extend the original model with multiple sub-models, referred to as ``experts''. Each expert specializes in distinct sub-tasks~\cite{riquelme2021scaling, fedus2022switch, zoph2022designing}. In each MoE layer, tokens from an input sequence first pass through a gating network, which determines their assignment to different experts.

\noindent\textbf{Federated Learning.} Federated learning~\cite{mcmahan2017communication} addresses privacy concerns by enabling multiple devices (i.e., participants) to cooperatively train a global model without exposing their local data. Specifically, devices share only model updates instead of raw data, thus decentralizing model training across devices~\cite{li2021survey, wei2020federated, chen2021fedgraph}. Due to its strong privacy guarantees, federated learning has been widely adopted in NLP model training~\cite{cai2023efficient, cai2023federated, wu2024fedbiot, kuang2024federatedscope}. In this paper, we focus on MoE fine-tuning in a federated setting, aiming to accelerate federated MoE fine-tuning convergence under resource constraints, which is still challenging in current federated NLP efforts due to the unique characteristics of MoE models.



\subsection{Motivation}
\label{sec:motivation}
The design of \textsc{Flux} is motivated by several important observations that are elaborated as follows.



\begin{figure}[t]
  \begin{minipage}[t]{.6\linewidth}  
    \centering
    \resizebox{\linewidth}{!}{
    \begin{tabular}{cccc} 
        \hline
        MoE-based LLM & \#L/\#E & \#Para. & Size \\
        \hline
        LLaMA-MoE~\cite{zhu2024llama} & 32/16 & 6.7B & 13.48GB\\
        Deepseek-MoE~\cite{dai2024deepseekmoe} & 28/64 & 16.4B & 32.77GB\\
        Deepseek-v2-lite~\cite{liu2024deepseek} & 27/64 & 15.7B & 31.44GB\\
        Mixtral-8x7B~\cite{jiang2024mixtral} & 64/8 & 46.7B & 96.82GB\\
        Qwen2-MoE~\cite{team2024qwen2} & 28/64 & 57.4B & 112.4GB\\
        \hline
    \end{tabular}
    }
    \captionof{table}{\label{tab:moes}MoE-based LLMs.}
    \label{tab:moes}
  \end{minipage}\hfill
  \begin{minipage}[t]{.36\linewidth}  
    \centering
    \vspace{-11mm}  
    \includegraphics[width=\linewidth]{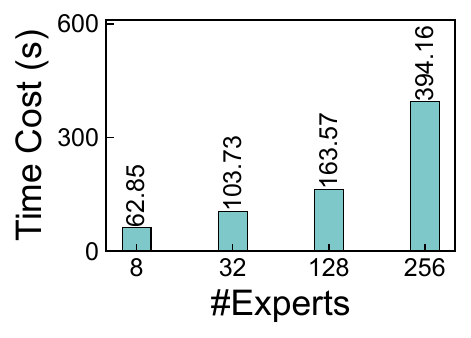}
    \vspace{-8mm} 
    \captionof{figure}{The fine-tuning costs.}
    \label{fig:cost_expert}
  \end{minipage}
\end{figure}

\subsubsection{Observation 1: MoE Fine-tuning is costly}
Typically, MoE-based LLMs have enormous model sizes, posing significant challenges for fine-tuning. As shown in Table \ref{tab:moes}, for example, the recently released LLaMA-MoE~\cite{llama-moe} has 32 layers with 16 experts each, requiring 13.48GB of memory just to store its parameters. Recent works propose updating only the expert parameters to improve efficiency, since the model performance relies heavily on experts~\cite{zadouri2023pushing, liu2024perft, wang2024let}. Nonetheless, fine-tuning remains costly even under expert-only updates, as expert parameters often account for more than two-thirds of the overall MoE model~\cite{zhang2023emergent, singh2023hybrid}. As shown in \autoref{fig:cost_expert}, we measure the one-round fine-tuning cost with different numbers of experts using the LLaMA-MoE model and 60 data samples from the Dolly dataset on a NVIDIA L20 GPU. The results show that as the number of experts increases, the fine-tuning cost grows significantly. 
The commonly used lossless memory optimization methods, such as gradient accumulation~\cite{gao2021scaling} and activation checkpointing~\cite{chen2016training, kirisame2020dynamic}, cannot address the unique challenge of MoE fine-tuning, as they do not reduce the size of the MoE model. Resource-constrained devices are therefore still unable to load the full model for fine-tuning. Although other techniques, such as quantization~\cite{kim2023mixture,frantar2023qmoe} and pruning~\cite{lu2024not,xu2024mome} can reduce expert size, they inevitably introduce computational errors, which can have adverse effects on fine-tuning convergence. Empirical evidence is provided in \S\ref{eva_result}.

\begin{figure}[t]
    \centering
    \includegraphics[width=\linewidth]{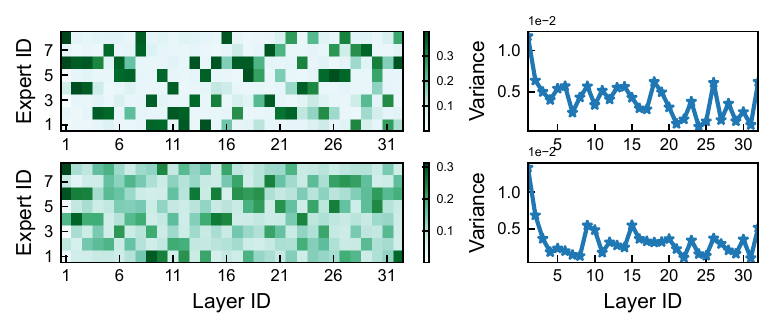}
    \caption{The activation frequencies (left) and corresponding variances (right) on experts.}
    \label{fig:activation_fre_var}
\end{figure}


\subsubsection{Observation 2: data utilization of experts varies}
Although all experts have the same number of parameters, their fine-tuning efficiency can vary due to differences in activation pattern across the training data. As shown in \autoref{fig:activation_fre_var}, we report expert activation frequencies on the GSM8K (upper) and MMLU (lower) datasets using the LLaMA-MoE model, where the activation frequency is calculated by dividing the number of activated tokens over the total number of tokens. 
Our findings reveal significant disparities in activation patterns. Some experts, such as expert-8 in the first layer, exhibit a high activation frequency, indicating high data utilization during fine-tuning. In contrast, other experts remain largely inactive. For instance, expert-3 in the first layer has an activation frequency of less than 5\%. This suggests that only 5\% of the data contribute to updating this expert's parameter updates, even if the full training dataset is processed. 


Moreover, we find that expert activation patterns vary significantly across different MoE layers. For instance, on the GSM8K dataset, the first layer exhibits a highly skewed activation distribution, where experts 4, 6, and 8 are frequently activated, while most others remain largely inactive. In contrast, the 31st layer demonstrates a more balanced activation pattern, with all experts being utilized at similar frequencies. A similar trend is observed in the MMLU dataset.  \autoref{fig:activation_fre_var} reports the variances of activation frequencies on different layers for a better illustration. When activation frequency is highly skewed, many experts receive minimal or no activation, reducing their contribution to the model’s predictions. Conversely, in layers with balanced activation, all experts participate more equally, suggesting a more distributed influence on the final output. These findings motivate the need for layer-specific expert merging strategies in \textsc{Flux} (\S\ref{merging strategy}). By tailoring the merging process to the activation characteristics of each layer, \textsc{Flux} can optimize resource utilization while maintaining model performance.

\begin{figure}[t]
\centering
        \subfigure[Fine-tuning convergence.]{
		\begin{minipage}[b]{0.215\textwidth}
			\includegraphics[width=1\textwidth]{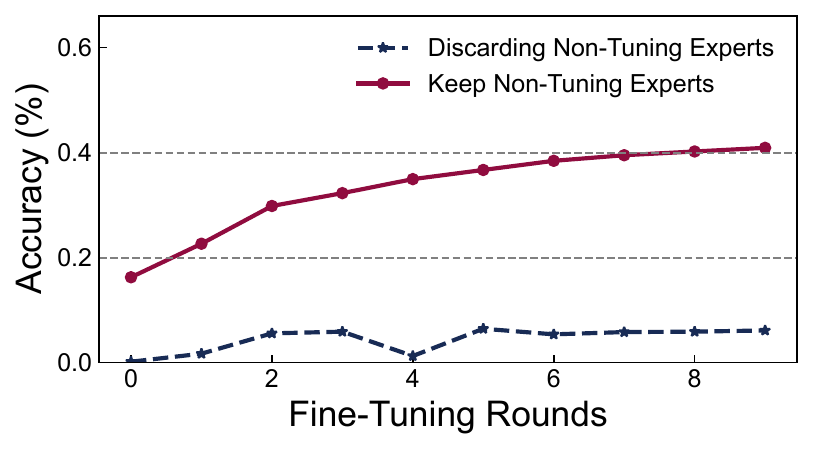} 
  \end{minipage}
		\label{fig:acc_drop}
	}
        \subfigure[Toy example.]{
            \begin{minipage}[b]{0.235\textwidth}
            \vspace{-10mm}
            \includegraphics[width=1\textwidth]{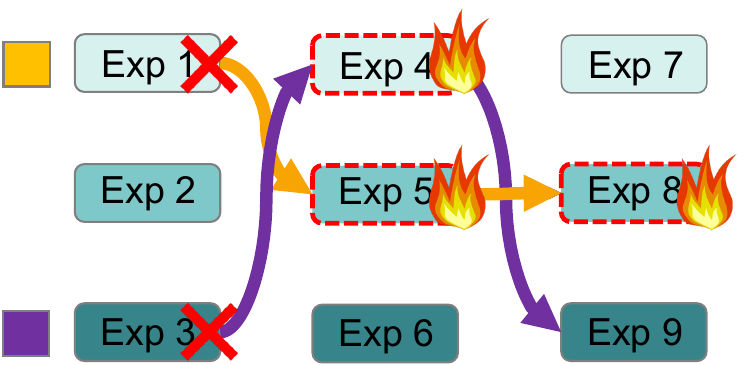}
            \end{minipage}
        \label{fig:merge_example}
        }
	\caption{The impact of discarding non-tuning experts.}
        \label{fig:expert_drop_moti}
\end{figure}

\subsubsection{Observation 3: non-tuning experts are also necessary} 
Given the variation in activation frequency across experts, a straightforward idea is to retain only experts with high activation frequency for fine-tuning while discarding the less activated ones, as proposed by FedMoE~\cite{mei2024fedmoe}. However, we argue that non-tuning experts are also essential for maintaining local fine-tuning performance. 
To evaluate their impact, we conduct experiments on the LLaMA-MoE model using the GSM8K dataset. Specifically, we compare two cases: preserving versus discarding non-tuning experts. In both cases, we fine-tune the 64 most frequently activated experts, following the approach in ~\cite{mei2024fedmoe}. We assess model performance using the ROUGE metric, a widely used measure of text generation quality, over 10 rounds of fine-tuning. As shown in \autoref{fig:acc_drop}, our results indicate that discarding non-tuning experts significantly degrades fine-tuning performance, leading to lower ROUGE scores.
These findings highlight the critical role of non-tuning experts in stabilizing training and maintaining overall model performance.

We further analyze the underlying rationale. Suppose we use the top-1 gating scheme, each token follows a specific forwarding path through different layers, activating selected experts along the way, as shown in \autoref{fig:merge_example}. Importantly, before reaching a highly activated expert, a token may first pass through one or more non-tuning experts. If these non-tuning experts are discarded, a mechanism must compensate for their absence, such as skipping expert computation at the affected layer or re-routing the token to other experts. However, these adjustments inevitably introduce errors into the model's computations. Worse still, these errors can propagate through subsequent layers, accumulating and ultimately hindering fine-tuning convergence. Therefore, it is necessary to develop a method that retains the essential information of non-tuning experts while controlling memory consumption, ensuring efficient and accurate MoE fine-tuning.


\subsubsection{Observation 4: expert role assignment is challenging}
To ensure the efficiency and effectiveness of federated MoE fine-tuning, it is crucial to determine which experts should be tuning or non-tuning, a problem referred to as expert role assignment. Intuitively, experts with a significant contribution to global fine-tuning convergence should be selected as tuning ones. However, expert role assignment is non-trivial due to three key challenges. First, a criterion should be defined to accurately measure expert contribution to the global fine-tuning convergence. A straightforward approach is to estimate an expert’s contribution based on its activation frequency~\cite{mei2024fedmoe}. While easy to implement, our experimental results show that this criterion fails to accurately capture each expert’s impact on global MoE convergence. Second, assessing the contribution of all experts on a participant is impractical due to limited memory capacity. However, without full knowledge of expert contributions, solving the expert role assignment problem accurately becomes challenging. 
Third, expert role assignment should account for system efficiency under participant computation heterogeneity. Greedily selecting the maximum number of experts to maximize total contribution for each participant may result in weaker participants experiencing prolonged fine-tuning times, thereby increasing the overall fine-tuning costs.
\section{System Overview}
\begin{figure}[t]
    \centering
    \includegraphics[width=\linewidth]{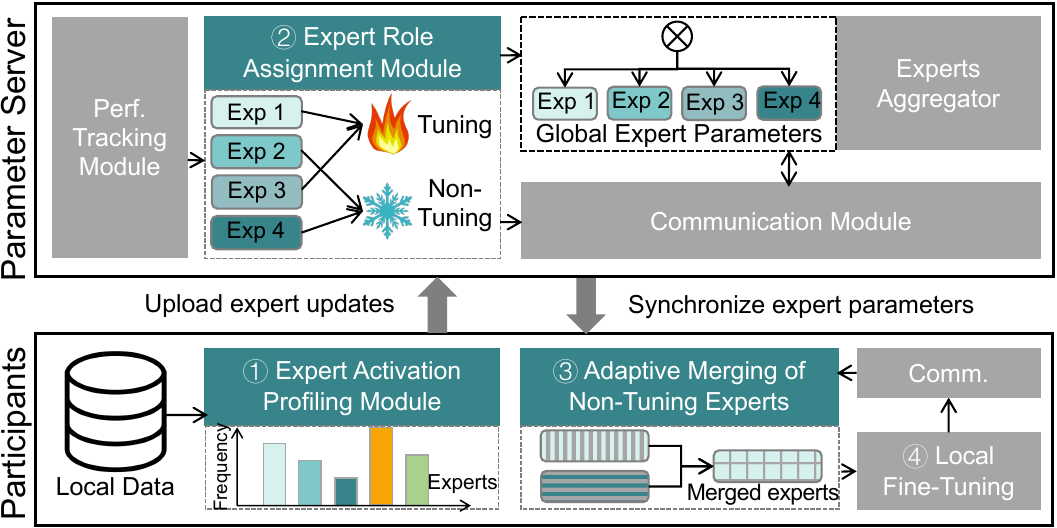}
    \caption{System overview of \textsc{Flux}.}
    \label{fig:system_overview}
\end{figure}


\autoref{fig:system_overview} illustrates the overall design of \textsc{Flux}. It is implemented using a classical parameter-server–based federated learning architecture~\cite{lai2021oort,cai2023federated,cai2023efficient,mei2024fedmoe}, where a central server coordinates a set of participants $N$ to collaboratively fine-tune the experts $E$ of an MoE-based LLM. Each participant fine-tunes the MoE model on its local dataset and uploads only model parameters, thereby avoiding exposure of local training data. Other methods, such as differential privacy~\cite{wei2020federated, wei2023personalized} and homomorphic encryption~\cite{hu2024maskcrypt, yan2024efficient}, are orthogonal to this work but can be incorporated into \textsc{Flux} to further enhance the privacy preservation during expert aggregation.
Due to memory constraints, each participant $i\in N$ can load a maximum of $B_{i}$ experts into GPUs for fine-tuning. In addition, to strictly control the round time, the maximum number of tuning experts at each participant $i$ is constrained to $B_{i}^{tune}$ based on local computing power, and the remaining $B_{i}^{non}=B_{i}-B_{i}^{tune}$ experts are frozen. With the goal of minimizing time-to-accuracy, \textsc{Flux} introduces three core modules:

\textbf{Expert Activation Profiling (\S\ref{sec:profiling})}: This module utilizes a quantized MoE model to profile expert activation, which is crucial for the local training of tuning experts and the merging of non-tuning experts. Since running a full MoE model for profiling is computationally infeasible, \textsc{Flux} performs profiling on a quantized version that closely mirrors the expert activation patterns of the full model. To further reduce overhead, \textsc{Flux} introduces a stale profiling scheme, enabling profiling to run concurrently with parameter aggregation, effectively hiding its time cost.

\textbf{Non-tuning Expert Merging (\S\ref{sec:merging})}: Unlike traditional methods that discard non-tuning experts, which negatively impact model performance, \textsc{Flux} employs an adaptive merging module to merge non-tuning experts, preserving model performance. Considering the varying activation patterns and merging effects across layers, \textsc{Flux} utilizes an adaptive expert layer size tailored for different layers. In addition, \textsc{Flux} adopts a similarity-based expert clustering approach and an importance-based merging strategy to further enhance model performance with merged non-tuning experts.

\textbf{Expert Role Assignment (\S\ref{sec:assignment})}: This module runs an expert role assignment algorithm to determine which experts are tuning or non-tuning for each participant. To accurately measure the contribution of experts to global fine-tuning convergence, \textsc{Flux} defines an expert utility metric based on gradients as a criterion for expert role assignment. To address the challenge of measuring expert utility under resource constraints, \textsc{Flux} dynamically explores the potential contribution of unselected experts. Furthermore, an efficient gradient estimation technique and a dynamic exploration-exploitation strategy are introduced to enhance the efficiency and effectiveness of the expert role assignment.

In addition to these core modules, \textsc{Flux} incorporates additional components, e.g., local training, communication, parameter aggregation and performance tracking, to support federated MoE fine-tuning. 
\section{Expert Activation Profiling}\label{sec:profiling}
This section details the use of a quantized model for efficient local profiling and explains how to further minimize its time cost by running profiling concurrently with parameter aggregation.


\begin{figure}[t]
    \centering
    \includegraphics[width=\linewidth]{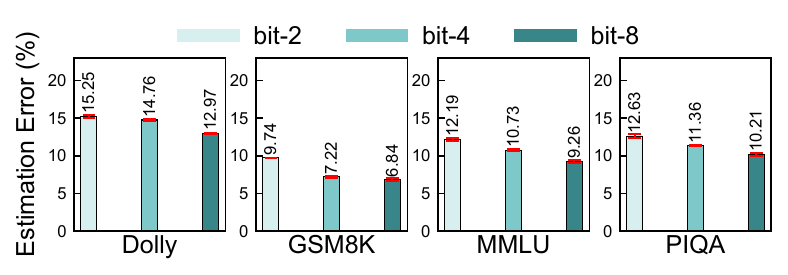}
    \caption{The Estimation error of activation frequency.}
    \label{fig:est_err}
\end{figure}

\subsection{Quantization-based Local Profiling}

Accurately profiling expert activation is a significant challenge in federated MoE fine-tuning. A straightforward approach is to run inference on local data using the full MoE model and track expert activations. However, this is often impractical for resource-constrained participants due to its high computational demands, making it either infeasible or prohibitively slow.
An alternative approach is to offload some data to a cloud server with sufficient computational resources, where expert activation can be directly measured using the full MoE model. However, this approach is incompatible with the federated setting that training data cannot be shared with others.

To address these challenges while ensuring both resource efficiency and data privacy, 
we propose to use a quantized MoE model to identify relevant data of each expert and estimate expert activation frequencies. Although quantized models cannot be directly used for fine-tuning, their activation patterns closely approximate those of the original full-precision models. This key observation allows us to leverage low-bit representations (e.g., INT4) to profile expert activation without requiring the full MoE model to run on resource-limited participants.

We evaluate the error in activation frequency estimation using different quantized MoE models, as shown in \autoref{fig:est_err}. 
Our results show that quantized MoE models can provide relatively accurate activation frequency estimations. For instance, using a 4-bit MoE model, the average estimation error is approximately 11.01\%. Furthermore, higher-precision models yield more accurate estimations, as their intermediate computations more closely match those of the original full-precision model. Since participants have varying computational resources, \textsc{Flux} allows each participant to flexibly choose the appropriate quantization level based on their available computing power, enabling efficient activation frequency estimation without overburdening local hardware.

By using the quantized MoE model, each participant $i$ can identify a subset of data $D_{i}^{e}$ that passes through each expert $e$. When a specific expert $e$ is selected for fine-tuning, we can feed the model with relevant datasets, thereby enhancing data utilization and fine-tuning efficiency.

\begin{figure}[t]
\centering
        \subfigure[Activation frequency changes over rounds.]{
		\begin{minipage}[b]{0.285\textwidth}
			\includegraphics[width=1\textwidth]{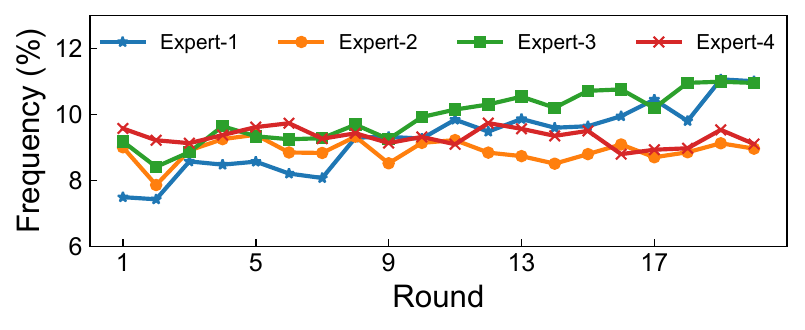} 
  \end{minipage}
		\label{fig:frq_change_round}
	}
        \subfigure[CDF of frequency change.]{
            \begin{minipage}[b]{0.165\textwidth}
            \includegraphics[width=1\textwidth]{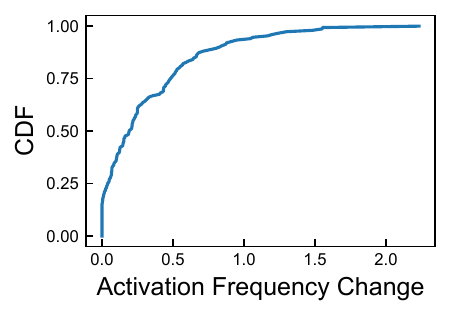}
            \end{minipage}
        \label{fig:frq_change_cdf}
        }
	\caption{Changes of activation frequency over rounds.}
        \label{fig:frq_change}
\end{figure}
\begin{figure}[t]
    \centering
    \includegraphics[width=\linewidth]{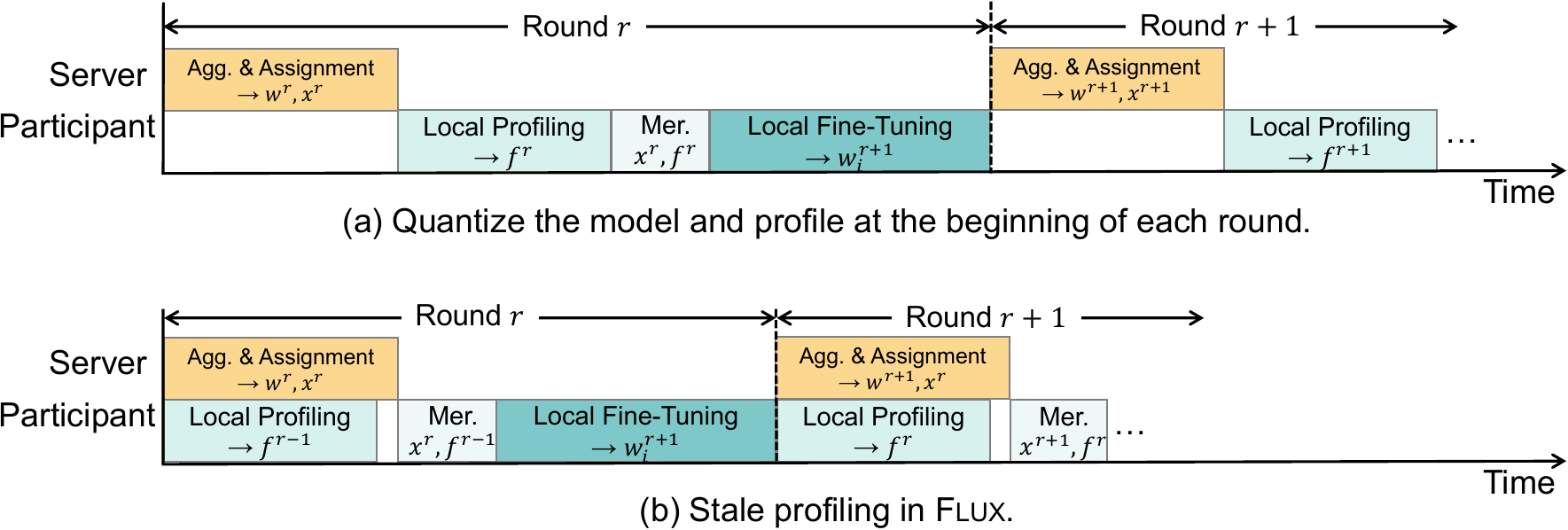}
    \caption{The delayed quantization and profiling mechanism, where $w^{r}$, $x^{r}$, and $f^{r}$ indicate the global MoE weight, expert role assignment decision, and estimated activation frequency at round $r$, respectively. }
    \label{fig:stale_quantization}
\end{figure}

\subsection{Stale Profiling}\label{stale quantization}

Expert activation pattern can shift over time due to parameter updates during the fine-tuning process. As shown in \autoref{fig:frq_change_round}, we track the activation frequencies of four experts in the LLaMA-MoE model on the GSM8K dataset over the first 20 fine-tuning rounds. For example, expert-1’s activation frequency starts at approximately 7.68\% but increases to 11.34\% after 20 rounds, demonstrating that initial estimation may become inaccurate over time. To maintain estimation accuracy, it is necessary to quantize and profile the latest MoE model downloaded from the parameter server in each round. 
However, frequent quantization and profiling introduce significant computational overhead, delaying the start of local fine-tuning and prolonging each fine-tuning round, as shown in~\autoref{fig:stale_quantization}(a).

To mitigate this overhead, \textsc{Flux} introduces a stale quantization and profiling mechanism, motivated by the observation that while expert activation frequencies do change, the difference between consecutive rounds remains small, as shown in~\autoref{fig:frq_change}. 
Therefore, we propose to conduct expert merging based on the stale profiling results from previous round, instead of waiting for quantization and profiling on the latest model $w^{r}$, as shown in
~\autoref{fig:stale_quantization}(b). Then, while the parameter server aggregates updates in round $r+1$, participants concurrently quantize and profile the local model. This parallel execution makes full use of the waiting time for model updates, significantly reducing delay and improving overall fine-tuning efficiency.

\section{Adaptive Merging of Non-Tuning Experts}\label{sec:merging}
Due to local resource constraints, each participant fine-tunes only a subset of experts. Discarding non-tuning experts can negatively impact performance, as discussed in the motivation section.
To address this challenge, \textsc{Flux} introduces an adaptive expert merging strategy, which retains the essential information of non-tuning experts while adhering to memory constraints. Prior studies~\cite{li2024merge} have demonstrated the feasibility of merging experts, but our approach faces additional challenges. 
First, given a total memory budget of $B_{i}^{non}$ for non-tuning experts, we must determine how much memory can be allocated for each layer. After that, we have to decide which experts should be merged and how to merge them in a way that minimizes information loss to preserve model performance.


\subsection{Adaptive Expert Layer Size}\label{merging allocation}
\begin{figure}[t]
\centering
        \subfigure[Dolly.]{
		\begin{minipage}[b]{0.225\textwidth}
			\includegraphics[width=1\textwidth]{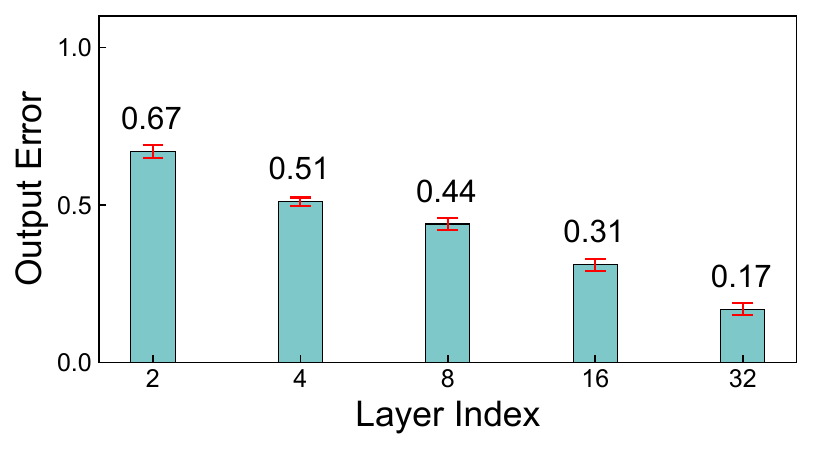} 
  \end{minipage}
		\label{fig:error_layer_dolly}
	}
        \subfigure[GSM8K.]{
            \begin{minipage}[b]{0.225\textwidth}
            \includegraphics[width=1\textwidth]{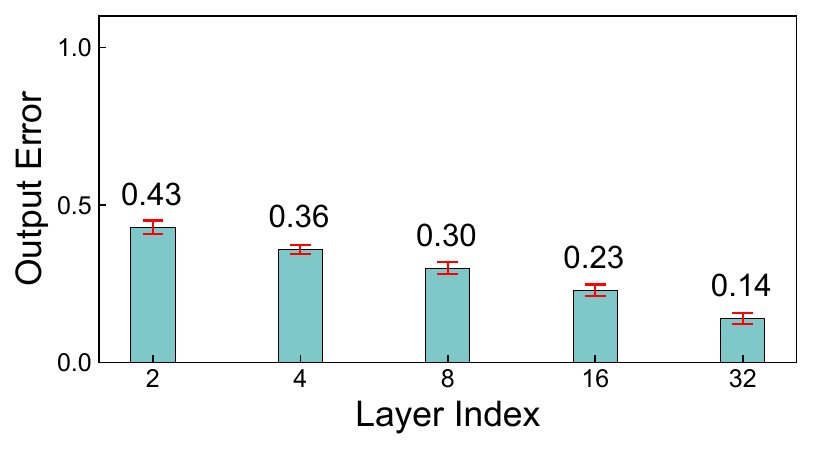}
            \end{minipage}
        \label{fig:error_layer_gsm8k}
        }
	\caption{Output error for merging in different layers.}
        \label{fig:error_layer}
\end{figure}

A naive approach would allocate an equal budget from $B_{i}^{non}$ for merged experts across all layers. However, our experimental results demonstrate that this uniform allocation is suboptimal (\S\ref{eva_abl}), as different layers exhibit distinct activation patterns and error accumulation characteristics. 

\textsc{Flux} dynamically determines the merging budget for each layer based on two key insights. (1) \textbf{Activation frequency distribution}: As shown in \autoref{fig:activation_fre_var}, expert activation patterns vary across different layers. In layers where activation frequency is highly skewed, some experts are rarely activated and can be merged with minimal impact. These layers require a smaller merging budget. Conversely, in layers with a more balanced activation distribution, all experts contribute meaningfully to the model output, making expert merging more detrimental to performance. These layers demand a larger merging budget to preserve expert diversity.
(2) \textbf{Error accumulation across layers}: As illustrated in \autoref{fig:error_layer}, we analyze the output error introduced by merging experts at different layers. The error is quantified as the average cosine distance between the final token embeddings from the MoE model with merged experts and those from the original full model. The results indicate that merging experts in earlier layers introduces significantly larger errors. This is because errors introduced during expert merging propagate forward through the network, accumulating and amplifying inaccuracies in deeper layers. Therefore, to minimize error accumulation, earlier layers should retain more expert information and thus be allocated a larger merging budget.

Based on these insights, \textsc{Flux} determines the merging budget for each layer $l$ as follows:

\begin{align}
    B_{i}^{non}(l) = \lfloor\frac{b_{i}^{l}}{\sum_{k=1}^{L} b_{i}^{l}} \times B_{i}^{non}\rfloor, \quad b_{i}^{l} = \frac{L-l+1}{v_{i}^{l}},
\end{align}
where $v_{i}^{l}$ is the variance of expert activation frequencies in layer $l$ for participant $i$. This adaptive strategy ensures that layers prone to higher error accumulation receive more resources, while layers with skewed activation distributions can effectively merge experts with minimal performance degradation. 

\subsection{Similarity-based Expert Clustering}\label{expert clustering}
After determining the merging budget for each layer, \textsc{Flux} proceeds to decide which experts should be merged together by formulating it as an expert clustering problem, i.e., clustering non-tuning experts in each layer $l$ into $B_{i}^{non}(l)$ groups.
\textsc{Flux} adopts a similarity-based merging strategy, motivated by the rationale that if two experts have similar parameters, their merged representation is more likely to preserve the original model’s behavior. Based on this insight, \textsc{Flux} follows a two-step clustering approach. First, to facilitate efficient clustering, \textsc{Flux} first reduces the dimensionality of expert parameters using standard techniques such as Principal Component Analysis (PCA)~\cite{wold1987principal}. The resulting low-dimensional representations serve as feature vectors for each expert, capturing their essential characteristics while reducing computational overhead.
Second, using these feature vectors, \textsc{Flux} applies the K-Means clustering algorithm to group non-tuning experts into $B_{i}^{non}(l)$ clusters per layer. Experts within the same cluster are merged, ensuring that the resulting merged experts retain as much information as possible from their individual components.

To improve efficiency, \textsc{Flux} fuses the expert clustering problems across layers, rather than handling each layer independently.
Specifically, we initialize $\sum_{l=1}^{L} B_{i}^{non}(l)$ centroids and label each with its corresponding layer index. 
To efficiently assign experts to centroids, \textsc{Flux} computes the cosine distances between all experts and all centroids using efficient matrix operations. To enforce layer-specific clustering, the distances between an expert and centroids from different layers are set to 0. Finally, experts are assigned to clusters following the standard K-Means algorithm, ensuring that the most similar experts are merged together while maintaining layer constraints.

\begin{figure}[t]
\centering
        \subfigure[Impact of discarding different experts on model outputs.]{
		\begin{minipage}[b]{0.22\textwidth}
			\includegraphics[width=1\textwidth]{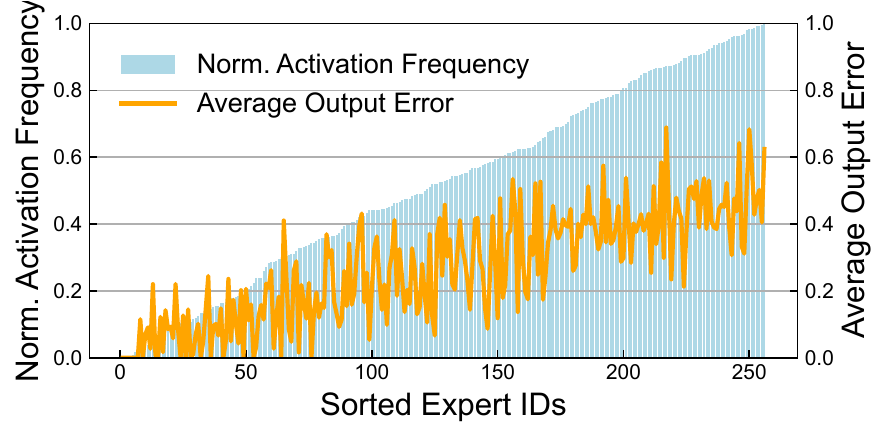} 
  \end{minipage}
		\label{fig:expert_sig}
	}
        \subfigure[The activation frequencies (\textcolor{darkblue}{$\blacksquare$}) and attention scores of experts (\textcolor{lightblue}{$\blacksquare$}).]{
            \begin{minipage}[b]{0.22\textwidth}
            \includegraphics[width=1\textwidth]{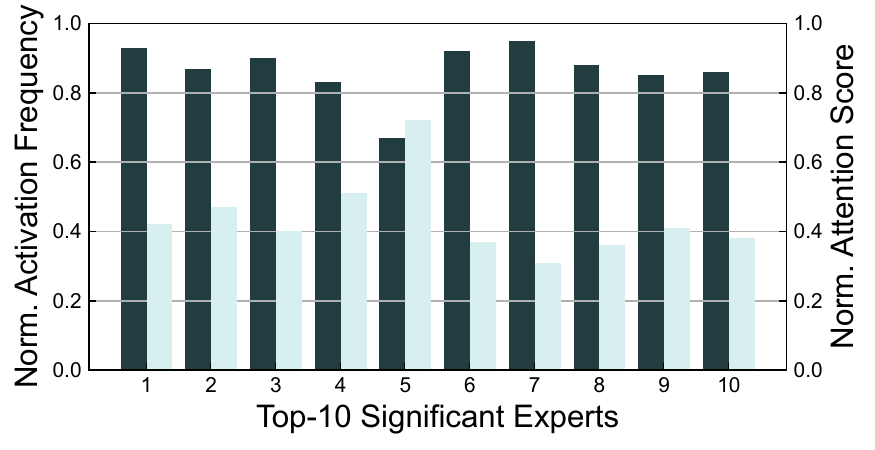}
            \end{minipage}
        \label{fig:expert_att}
        }
	\caption{Expert significance.}
\end{figure}

\subsection{Importance-based Merging Strategy}\label{merging strategy}
For experts within the same clustering group, \textsc{Flux} should determine how to merge them together to minimize the negative impact on the model output. 
A naive approach would be to merge experts based solely on their activation frequency, assuming that less frequently activated experts contribute less to model outputs~\cite{li2024merge}.
However, our analysis reveals that activation frequency alone is not a reliable indicator of expert significance. Some experts with low activation frequencies still play a crucial role in determining model outputs, making their preservation during merging essential.

To quantify the impact of different experts, we report the average output error when discarding experts using the LLaMA-MoE model. The output error is measured as the distance between the model's output with discarded experts and its output without discarded experts. As shown in \autoref{fig:expert_sig}, a larger output error indicates that an expert plays a more critical role and should be retained as much as possible during merging. Surprisingly, expert significance does not always correlate with activation frequency. This phenomenon can be explained as follows. Some experts specialize in processing a small subset of tokens. However, these tokens exhibit high attention scores to other tokens, meaning their outputs influence multiple token representations across different layers. Discarding such experts disrupts high-attention token processing, leading to incorrect or suboptimal model outputs.

To further validate this observation, we analyze the average attention scores of tokens processed by the top 10 most significant experts, as shown in \autoref{fig:expert_att}. Notably, the fifth expert, despite having a low activation frequency of just 0.67, ranks among the most significant experts. While many other experts exhibit higher activation frequencies, the tokens processed by this expert have exceptionally high attention scores, amplifying its influence on overall model outputs.

Based on the above insights, \textsc{Flux} merges non-tuning experts of each cluster $c$ by
\begin{align}
    W_{\text{merged}}=\sum_{e\in E_{c}}\frac{\alpha_{e}}{\sum_{e^{'}\in E_{c}}\alpha_{e^{'}}}W_{e},\quad\quad \alpha_{e} =f_{e}\cdot\bar{a}_{e}
\end{align} 
where $W_{e}$ represents the parameters of expert $e$, $E_{c}$ denote the set of non-tuning experts belonging to cluster $c$, and $\bar{a}_{e}$ is the average attention value of tokens assigned to expert $e$. Additionally, $f_{e}$ represents the activation frequency of expert $e$, which is estimated locally using the quantized MoE model.

\section{Dynamic Expert Role Assignment}\label{sec:assignment}

The expert role assignment module in Flux is responsible to choose a ``good'' subset of experts for fine-tuning in each round under resource constraints. The basic idea is to calculate a utility value for each expert, quantifying its contribution to fine-tuning convergence. Flux prioritizes experts with higher utility, aiming to reduce the number of training rounds needed for convergence. Simultaneously, the selection process enforces resource constraints on each participant, ensuring that the round duration remains controlled and does not exceed device capabilities. Before presenting the formal expert selection algorithm, we first outline several key design choices that guide the selection process. 

\subsection{Key Design Choices}
\textbf{Expert utility definition}. 
The first key design choice is the definition of expert utility, which should accurately reflect an expert's contribution to fine-tuning, particularly in the presence of non-IID data distributions.
Motivated by~\cite{katharopoulos2018not, lai2021oort}, Flux defines the utility of an expert $e$ for participant $i$ as follows.
\begin{align}
    u^{e}_{i} = |D_{i}^{e}|\sqrt{\frac{1}{|D_{i}^{e}|}\sum\limits_{k\in D_{i}^{e}}\nabla g_{k}},\label{utility}
\end{align}
where $D_{i}^{e}$ denotes the set of tokens passing through the expert $e$ on participant $i$, and $g_{k}$ represents the gradient of token $k\in D_{i}^{e}$. Note that $D_{i}^{e}$ could be obtained by the profiling module and $g_{k}$ comes from the previous training round. This definition ensures that expert utility is data-driven, capturing the actual contribution of an expert to model updates.

\noindent\textbf{Expert utility update}.
With the expert utility defined in (\ref{utility}), only the experts selected in the previous training round have their utility values refreshed, leaving unselected experts without updated estimates. 
To address this challenge, \textsc{Flux} employs an \textbf{exploration-exploitation} strategy: in each training round, a random subset of experts is selected to explore their potential utility. This allows the model to gradually refine utility estimates for all experts, even those that have not yet been selected. The balance between exploitation (choosing high-utility experts) and exploration (testing unselected experts) ensures that \textsc{Flux} makes informed selection decisions over time.

\noindent\textbf{Expert role assignment}.
After obtaining utility values for candidate experts, \textsc{Flux} selects tuning experts by solving the following optimization problem:
\begin{align}
    \max \quad&\sum_{i}\sum_{e}x_{i}^{e}u_{i}^{e}, \quad\text{s.t.,}
    & \sum_{e}x_{i}^{e}\leq B_{i}^{tune},\forall i\in N;\label{opt}
\end{align}
where $x_{i}^{e}$ is a binary variable indicating whether expert $e$ is selected for fine-tuning on participant $i$. The constraint enforces that the number of tuning experts does not exceed the available capacity $B_{i}^{tune}$. Note that $B_{i}^{tune}$ is determined by local computing resources and round duration constraint. 

\subsection{Algorithm Design}\label{expert selection alg}
Bringing together the previous components, we now formally present the expert selection algorithm in \autoref{alg:fed_alg}. At the beginning of each training round, the parameter server collects expert utility values from all participants. Note that in the first round, utility values of experts are initialized based on their activation frequencies, i.e., $u_{i}^{e}=\text{Norm}(a_{i}^{e})$. 

After collecting the utility values, the parameter server solves the optimization problem (\ref{opt}) to determine a set of candidate experts $E_{i}=\{e|\tilde{x}_{i}^{e}=1\}$ for each participant. However, rather than fine-tuning all experts in $E_{i}$, \textsc{Flux} balances exploitation and exploration as follows. A portion $\epsilon$ of experts from $E_{i}$ with the highest utility values is selected for exploitation. The remaining $(1-\epsilon)$ experts are randomly selected for exploration and they are included in set $E_{i}^{exl}$, where $|E_{i}^{exl}|=(1-\epsilon)|E_{i}|$. Finally, selection results ${E}_{i}^{exp}$ and $E_{i}^{exl}$ are sent to the corresponding participant $i$.

\begin{algorithm}[t]
\caption{Expert Role Assignment in \textsc{Flux}}
\label{alg:fed_alg}
\SetAlgoLined
Collect utilities of experts from participants\;
Solve optimization problem (\ref{opt}) to obtain solution $\tilde{x}_{i}^{e}$\;
\For{participant $i=1,2,...$}{
    Get a set of candidate experts as $E_{i}=\{e|\tilde{x}_{i}^{e}=1\}$\;
    Select a subset of tuning experts ${E}_{i}^{exp}\subseteq E_{i}$ with the highest utility values for exploitation, and $|{E}_{i}^{exp}|=\epsilon|E_{i}|$\;
    Randomly select a subset of experts $E_{i}^{exl}$ for exploration and $|E_{i}^{exl}|=(1-\epsilon)|E_{i}|$\;
    Send ${E}_{i}^{exp}$ and $E_{i}^{exl}$ to participant $i$\;
}
\end{algorithm}

\noindent\textbf{Efficient Gradient Estimation for $E_{i}^{exl}$. }
For exploitation experts, gradients are obtained directly through backpropagation during fine-tuning, allowing utility computation for expert selection in the next round. However, applying the same approach to exploration experts would be inefficient, as their contribution to fine-tuning is often trivial, leading to unnecessary computational overhead. Since we only need gradient estimates for exploration experts rather than full parameter updates, \textsc{Flux} adopts a forward-only gradient estimation method~\cite{baydin2022gradients, feng2024baffle}. Small perturbations, randomly sampled from a normal distribution, are added to the expert parameters. The expert gradient is estimated by averaging loss differences obtained with different perturbation values. 

This eliminates the need for backpropagation of exploration experts, significantly improving efficiency. Note that this forward-only apparoach may have estimation errors, and it is only applied to exploration experts, as accurate gradients are still required for exploitation experts to ensure precise parameter updates.

\noindent\textbf{Dynamic Exploration and Exploitation. } 
The trade-off between exploration and exploitation is governed by the hyperparameter 
$\epsilon$. A smaller $\epsilon$ prioritizes exploration, updating utilities for more experts but potentially slowing global convergence due to the inclusion of low-utility experts. Conversely, a larger $\epsilon$ favors exploitation, selecting experts based on the optimization solution, but may result in some experts being updated infrequently and limit adaptability.

To balance this trade-off, \textsc{Flux} adopts a dynamic exploration and exploitation strategy, where $\epsilon$ gradually increases as federated fine-tuning progresses. The rationale is that as more rounds are completed, utility estimations become more reliable, improving expert selection accuracy. This allows \textsc{Flux} to prioritize optimization-based selection over time, maximizing total utility while maintaining adaptability in earlier stages.

\section{Implementation}
We implement a prototype system of \textsc{Flux} with approximately 3K lines of code using Python and PyTorch. \textsc{Flux} manages the metadata of each participant through an isolated object, which includes information such as participant ID, activation frequency profiling, and other relevant attributes. \textsc{Flux} supports the integration of additional fine-tuning optimization techniques, such as Adapter~\cite{houlsby2019parameter} and LoRA~\cite{hu2022lora}. To realize the proposed optimizations, \textsc{Flux} incorporates the following specialized functions:



\noindent\textbf{Customized MoE construction. }\textsc{Flux} enables users to define customized MoE models with flexible expert scales in each layer, differing from existing frameworks that enforce an equal number of experts per MoE layer. To achieve this, \textsc{Flux} provides the API \texttt{Flux.moe.customized\_moe(model, exps\_config)}, which serves as an entrance for modifying the original MoE layer with \textsc{Flux}'s customized MoE layer. Here, \texttt{exps\_config} is a list or dictionary specifying the number of experts per MoE layer.  

\noindent\textbf{Pre-trained model loading. }With a customized MoE model, the standard model parameters cannot be directly loaded from the original checkpoint file due to architectural modifications. \textsc{Flux} provides the API \texttt{Flux.moe.load\_model(\allowbreak model\_path,\allowbreak exps\_config)}, allowing users to load parameters for the customized MoE. This API encapsulates the \texttt{\allowbreak transformers.\allowbreak AutoModel.\allowbreak from\_pretrained} function, and loads parameters for experts and other components (e.g., attn) separately.  

\noindent\textbf{Gate re-routing. }After expert merging, the gating mechanism must be updated to ensure that tokens are correctly routed to merged experts instead of the original ones. To this end, \textsc{Flux} remaps the original experts assigned by the gate network to new destinations if they have been merged.

\section{Performance Evaluation}
\subsection{Setup}

\begin{figure*}[t]
    \centering
    \includegraphics[width=\linewidth]{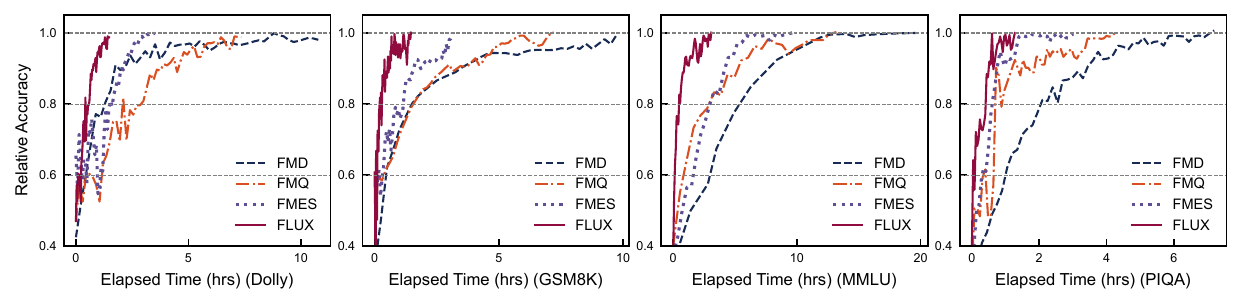}
    \caption{Convergence throughout the fine-tuning process on the LLaMA-MoE model.}
    \label{fig:convergence_all_llama}
\end{figure*}

\begin{figure*}[t]
    \centering
    \includegraphics[width=\linewidth]{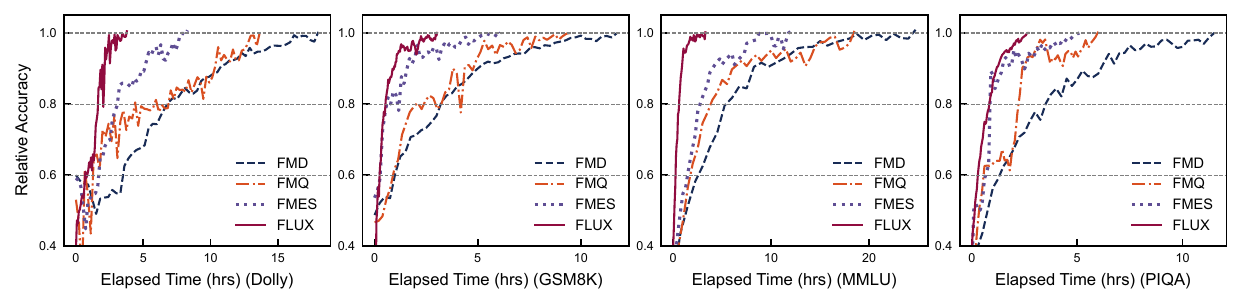}
    \caption{Convergence throughout the fine-tuning process on the DeepSeek-MoE model.}
    \label{fig:convergence_all_deepseek}
\end{figure*}

\noindent\textbf{Environment. }
We evaluate \textsc{Flux} on a testbed consisting of NVIDIA L20 GPUs with 48GB GPU memory, interconnected via PCIe. The system runs Ubuntu 20.04 with Linux kernel version 5.15.0, NVIDIA driver 550.67, and CUDA 11.8. 
We use PyTorch 2.6.0 and transformers 4.49.0 for model fine-tuning.

\noindent\textbf{Model and Datasets.} We evaluate the performance of \textsc{Flux} using the LLaMA-MoE~\cite{llama-moe} and DeepSeek-MoE~\cite{dai2024deepseekmoe} models, with details summarized in Table \ref{tab:moes}. We evaluate these models on four diverse datasets: (1) Dolly~\cite{DatabricksBlog2023DollyV2} is an open dataset containing over 15K records, generated by thousands of Databricks employees. (2) GSM8K~\cite{cobbe2021gsm8k} consists of 8.5K high-quality, linguistically diverse grade-school math problems. (3) MMLU~\cite{hendryckstest2021} is a massive multitask benchmark comprising multiple-choice questions spanning a broad range of knowledge domains. (4) PIQA~\cite{Bisk2020} is a dataset for commonsense reasoning, designed to evaluate a model’s understanding of physical knowledge in NLP. 
All datasets are partitioned into non-IID subsets following the FedNLP benchmark~\cite{lin2022fednlp}, reflecting realistic data heterogeneity across participants.

\noindent\textbf{Baselines. }We compare \textsc{Flux} against the following baselines. 
(1) \textit{Federated MoE fine-tuning with offloading (FMD)}: 
This is the default approach for fine-tuning MoE models under resource constraints, where inactive experts are dynamically offloaded to the CPU to enable fine-tuning within limited GPU memory. This method has been widely used in MoE inference under resource constraints~\cite{eliseev2023fast,hwang2024pre}.
(2) \textit{Federated MoE fine-tuning with quantization (FMQ)}: This approach quantizes all expert parameters from FP32 to INT4, allowing each participant to fit the entire MoE model into local memory for fine-tuning~\cite{lee2024owq,zhang2024quantized}. 
(3) \textit{Federated MoE fine-tuning with expert selection (FMES)}: This approach, also adopted in~\cite{mei2024fedmoe}, selects a subset of experts for fine-tuning while considering resource constraints. Expert selection is based on activation frequencies, prioritizing frequently activated experts for fine-tuning. 

\noindent\textbf{Metrics. }
We primarily evaluate performance using the time-to-accuracy metric, a standard approach in federated learning studies~\cite{cai2023efficient,cai2023federated}. Each dataset is split into 80\% for fine-tuning and 20\% for testing. 
Since different datasets are usually with distinct evaluation metrics, we set dataset-specific target values and report the elapsed time required to reach these targets. The target values for each dataset are set as follows: 0.5 (ROUGE-L) for Dolly~\cite{qin2024federated}; 0.62 (Accuracy) for GSM8K~\cite{liu2024gold}; 0.75 (Accuracy) for MMLU~\cite{vavre2024llama}; and 0.8 (Accuracy) for PIQA~\cite{li2024evaluating}. We report the relative accuracy in our experiments (e.g., \autoref{fig:convergence_all_llama} and \autoref{fig:convergence_all_deepseek}), which is defined as the ratio between the obtained evaluation score (e.g., ROUGE-L and Accuracy) and the corresponding target value. 
We also report other metrics, e.g., activation frequency estimation error and algorithm time costs, for ablation study.

\noindent\textbf{Other settings.} 
\textsc{Flux} and all baseline methods are evaluated using the same set of hyper-parameters to ensure a fair comparison. Specifically, we set the mini-batch size to 16, the local training iterations per round to 1, and the learning rate to 1e-5. We set the number of participants per round to 20.
For all methods, the parameter server aggregates expert parameters using the FedAvg strategy~\cite{mcmahan2017communication}.

\subsection{Overall Performance} \label{eva_result}
\textbf{Convergence}. We first evaluate the convergence of different methods under a 10-participant setting, with results presented in \autoref{fig:convergence_all_llama} and \autoref{fig:convergence_all_deepseek}.
FMQ exhibits unstable convergence across all datasets, primarily due to quantization-induced errors in backpropagation. These errors lead to suboptimal parameter updates, preventing the model from converging effectively.
FMD achieves stable convergence since it involves all experts in both MoE forward and backward propagation. However, its reliance on frequent expert offloading between CPU and GPU results in high I/O communication overhead, significantly prolonging fine-tuning time due to resource constraints.
\textsc{Flux} outperforms FMES due to its sophisticated expert selection strategy and adaptive merging of non-tuning experts. Unlike FMES, which relies solely on activation frequency for expert selection, \textsc{Flux} dynamically selects high-utility experts while merging non-tuning experts, leading to faster and more stable convergence.

We also report the final ROUGE-L and accuracy values in \autoref{tab:break_down}. Both FMQ and FMES compromise fine-tuning quality, as they either compress experts or drop non-tuning experts, which damages model outputs and negatively affects fine-tuning results. In contrast, \textsc{Flux} largely preserves fine-tuning accuracy, with only a small gap compared to full MoE fine-tuning (FMD).

\begin{figure*}[t]
    \centering
    \includegraphics[width=\linewidth]{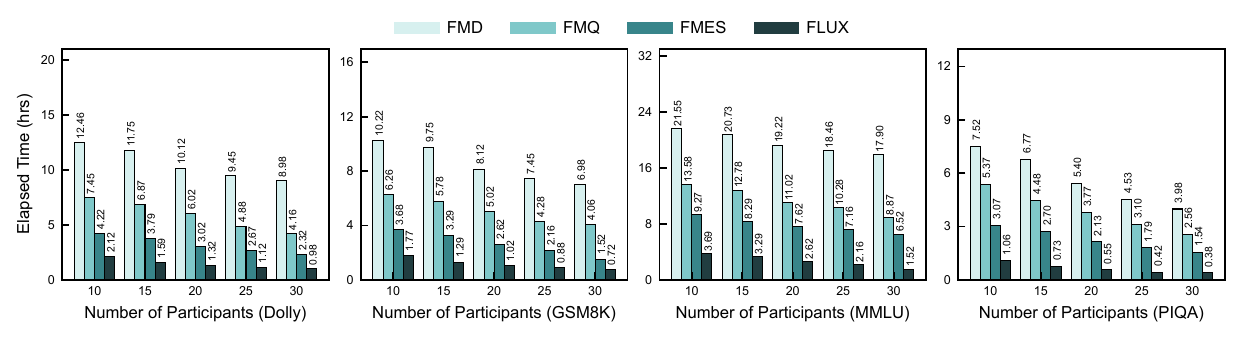}
    \caption{Time-to-accuracy with different numbers of participants on the LLaMA-MoE model.}
    \label{fig:scalability_all_llama}
\end{figure*}

\begin{figure*}[t]
    \centering
    \includegraphics[width=\linewidth]{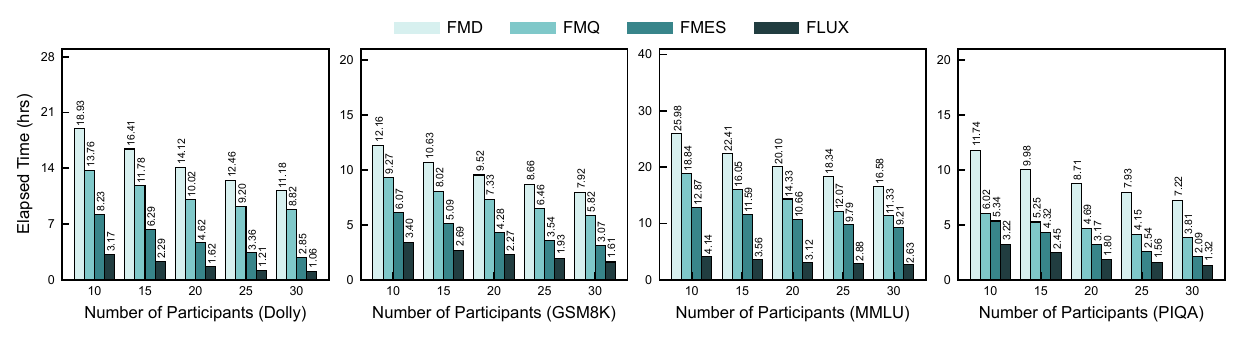}
    \caption{Time-to-accuracy with different numbers of participants on the DeepSeek-MoE model.}
    \label{fig:scalability_all_deepseek}
\end{figure*}

\noindent\textbf{Scalability}. To evaluate the scalability of \textsc{Flux}, we measure its performance under different numbers of participants, varying from 10 to 30, and record the elapsed time required to reach the target accuracy. As shown in \autoref{fig:scalability_all_llama} and \autoref{fig:scalability_all_deepseek}, \textsc{Flux} significantly accelerates federated MoE fine-tuning convergence, achieving a performance improvement of approximately 5.36$\times$ on LLaMA-MoE and 4.14$\times$ on DeepSeek-MoE compared to baseline approaches. As the number of participants increases, the time required to reach the target accuracy decreases, benefiting from greater computational parallelism in the federated fine-tuning process. However, the rate of improvement gradually diminishes as more participants join, due to the increased communication overhead associated with model updates and synchronization.

\noindent\textbf{Impact of datasets}. 
The acceleration provided by \textsc{Flux} varies across different datasets. For instance, \textsc{Flux} accelerates federated MoE convergence by approximately 4.91$\times$ on the GSM8K dataset with the LLaMA-MoE model, whereas the acceleration increases to 5.18$\times$ on the MMLU dataset, as shown in \autoref{fig:scalability_all_llama}. 
In addition, we observe that DeepSeek-MoE requires longer fine-tuning time than LLaMA-MoE across all datasets due to its larger model size. The increase in fine-tuning time is most pronounced on the Dolly dataset. 
In contrast, GSM8K exhibits a smaller increase in fine-tuning time across different models. This is primarily due to differences in sequence length among dataset samples. 
\begin{table}[t]
\centering
\resizebox{\linewidth}{!}{
\begin{tabular}{c|c|c|c|c|c}
\hline
\textbf{Model} & \textbf{Method} & \textbf{\makecell[c]{Dolly\\(ROUGE-L)}} & \textbf{\makecell[c]{GSM8K\\(Accuracy)}}\ & \textbf{\makecell[c]{MMLU\\(Accuracy)}} & \textbf{\makecell[c]{PIQA\\(Accuracy)}}\\
\hline
\multirow{4}{*}{\makecell[c]{LLaMA-\\MoE}} & \textbf{FMD} & 0.528 & 0.665 & 0.795 & 0.849 \\
~ & \textbf{FMQ} & 0.504 & 0.614 & 0.759 & 0.802 \\
~ & \textbf{FMES} & 0.518 & 0.622 & 0.774 & 0.826 \\
~ & \textbf{FLUX} & 0.527 & 0.663 & 0.793 & 0.848 \\
\hline
\multirow{4}{*}{\makecell[c]{DeepSeek-\\MoE}} & \textbf{FMD} & 0.529 & 0.669 & 0.801 & 0.853 \\
~ & \textbf{FMQ} & 0.507 & 0.618 & 0.765 & 0.805 \\
~ & \textbf{FMES} & 0.519 & 0.625 & 0.775 & 0.830 \\
~ & \textbf{FLUX} & 0.529 & 0.665 & 0.798 & 0.851 \\
\hline
\end{tabular}
}
\caption{Final achieved ROUGE and accuracy values using different methods.}
\label{tab:break_down}
\end{table}

\begin{figure}[t]
    \centering
    \includegraphics[width=\linewidth]{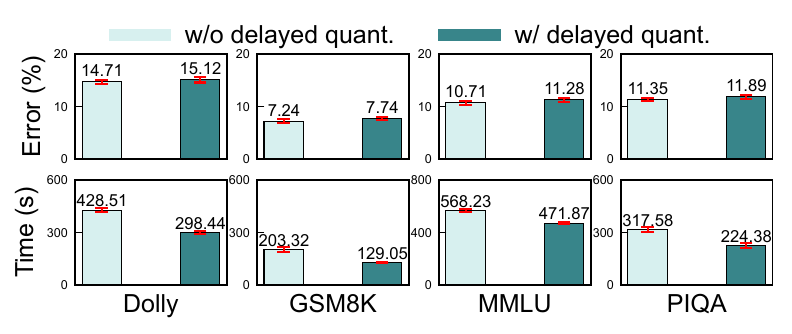}
    \caption{The impact of stale profiling.}
    \label{fig:quant_delay}
\end{figure}

\subsection{Ablation Study}\label{eva_abl}

\noindent\textbf{Effectiveness of stale profiling}.
We first study the impact of stale profiling proposed in \S\ref{stale quantization}, by comparing the activation frequency estimation errors with and without stale optimization mechanism. We use 2-bit quantized model for profiling. 
As shown in \autoref{fig:quant_delay}, we can see that stale profiling brings a negligible growth, less than 2\%, in estimation error. On the other hand, stale quantization significantly reduces the fine-tuning round time, decreasing it by approximately 28.2\%. This improvement is attributed to the ability to start local fine-tuning earlier while executing the quantization and profiling operations in parallel with the expert selection process.

\begin{figure}[t]
    \centering
    \includegraphics[width=\linewidth]{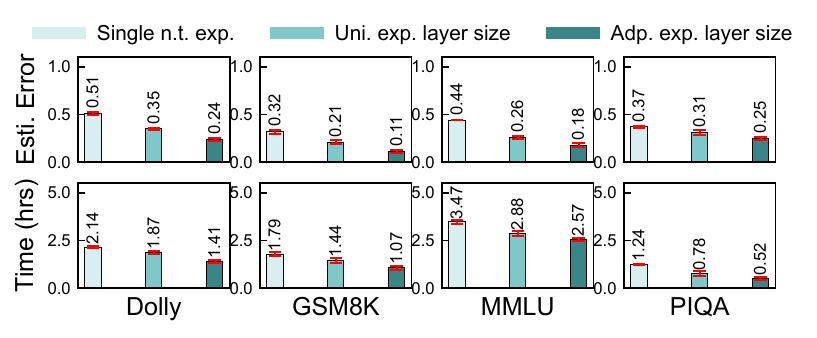}
    \caption{Impact of adaptive expert layer size.}
    \label{fig:merging_selecting}
\end{figure}

\noindent\textbf{Impact of adaptive expert layer size}.
We next evaluate the impact of adaptive expert layer size allocation (\S\ref{merging allocation}) by comparing \textsc{Flux} with two baselines. (1) Single non-tuning expert: merges all non-tuning experts within the same layer into a single expert without considering layer-wise merging budgets. (2) Uniform layer size: distributes the total merging budget evenly across all layers, disregarding variations in expert importance and activation patterns. We show the elapsed time to reach the target accuracy and forward-pass output errors in \autoref{fig:merging_selecting}. Note that forward-pass output errors are defined as the average cosine distance between token outputs of the merged-expert MoE model and the original MoE model. 
\textsc{Flux}’s adaptive merging strategy significantly reduces output error compared to the baselines. On the GSM8K dataset, \textsc{Flux} reduces output error by 65.6\% and 47.6\% relative to the Single Non-Tuning Expert and Uniform Layer Size baselines, respectively. 

\begin{figure}[t]
    \centering
    \includegraphics[width=\linewidth]{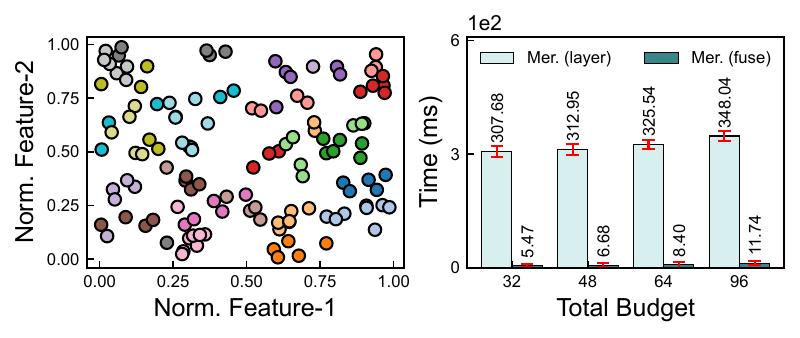}
    \caption{Cost of clustering 128 non-tuning experts.}
    \label{fig:k-mean}
\end{figure}

\noindent\textbf{Efficiency of expert clustering}.
We then evaluate the efficiency of expert clustering proposed in (\S\ref{expert clustering}) by first visualizing clustering results in \autoref{fig:k-mean}, where experts in the same cluster are denoted by the same color. We can clearly see that similar non-tuning experts are clustered together. To further analyze the impact of cross-layer clustering fusion, we compare it against a layer-wise independent clustering approach, where expert clustering is conducted separately for each layer. Our findings show that fusing the clustering process across all layers significantly reduces computation time. Specifically, independent clustering for each layer incurs an average time cost of 323.55 ms. Cross-layer clustering fusion reduces this time to 8.07 ms, achieving a 40× speedup. 
This substantial improvement is primarily attributed to the elimination of redundant operations, such as repeated centroid initialization across layers.


\begin{figure}[t]
    \centering
    \includegraphics[width=\linewidth]{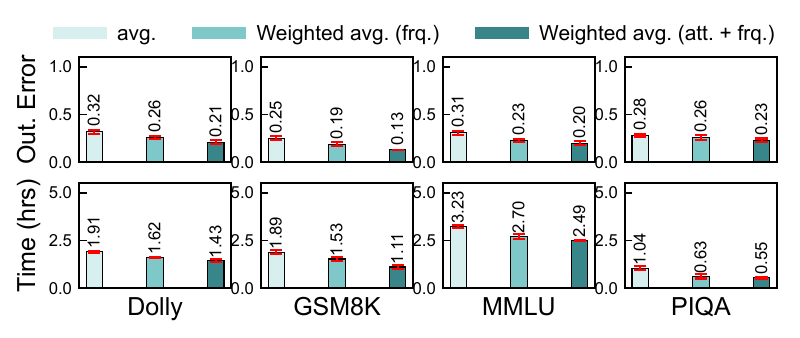}
    \caption{Efficiency of different merging strategies.}
    \label{fig:merging_method}
\end{figure}

\noindent\textbf{Efficiency of merging strategy}.
We further evaluate the efficiency of the expert merging strategy proposed in \S\ref{merging strategy} by comparing it with two baseline methods: averaging-based merging (Avg.) that merges experts by directly averaging their parameters, and activation-frequency-based weighted merging (Weighted Mer. (Frq.)) that merges experts using weights proportional to their activation frequencies~\cite{li2024merge}. \textsc{Flux} adopts a more refined approach, Weighted Mer. (Att. + Frq.), which considers both activation frequencies and token attention on experts during merging. We configure the expert layer size and clustering settings identically across all methods to ensure a fair comparison.
Output errors and elapsed time to target accuracy are shown in \autoref{fig:merging_method}. \textsc{Flux}’s merging strategy effectively minimizes output error. For example, on the Dolly dataset, \textsc{Flux} reduces output error by an additional 34.4\% and 19.2\% compared to Avg. and Weighted Mer. (Frq.), respectively. The output error on the GSM8K dataset is generally lower due to shorter sentence lengths, which naturally result in less accumulated merging error. With lower output error, \textsc{Flux} achieves a significant speedup in fine-tuning convergence, completing the training process 1.37× faster than the baseline methods.


\begin{figure}[t]
    \centering
    \includegraphics[width=\linewidth]{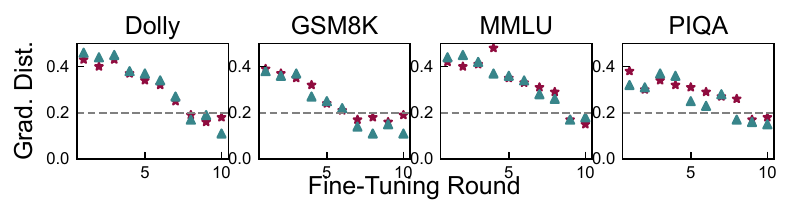}
    \caption{The effectiveness of the gradient estimation.}
    \label{fig:grad_estimation}
\end{figure}

\noindent\textbf{Effectiveness of gradient estimation}.
The effectiveness of gradient estimation, introduced in \S\ref{expert selection alg} for fast gradient estimation of non-tuning experts, is evaluated in~\autoref{fig:grad_estimation}, where we report the average normalized cosine distance between the estimated gradient and the ground truth over 10 consecutive fine-tuning rounds. Note that the ground truth is obtained through back-propagation. We observe that \textsc{Flux}’s gradient estimation method accurately approximates true gradients, with an average absolute distance of 0.29. Moreover, we observe that this distance gradually decreases as fine-tuning progresses. This trend may be attributed to the natural decline in gradient magnitudes as the model converges, which results in more stable and precise gradient estimations over time.

\begin{figure}[t]
    \centering
    \includegraphics[width=\linewidth]{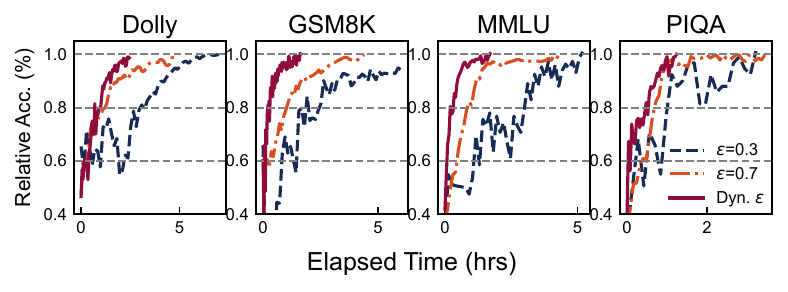}
    \caption{Performance of different $\epsilon$ strategies.}
    \label{fig:epsilon}
\end{figure}

\noindent\textbf{Impact of dynamic $\epsilon$}.
We study the impact of dynamic $\epsilon$ in exploration-exploitation strategy (\S\ref{expert selection alg}) by comparing it with two fixed strategies: $\epsilon=0.3$ and $\epsilon=0.7$. We report the convergence rate and elapsed time to target accuracy in \autoref{fig:epsilon}. Our results demonstrate that the dynamic $\epsilon$ significantly accelerates convergence by dynamically adjusting the balance between exploring expert utilities and optimizing global fine-tuning convergence. Fixed $\epsilon=0.3$ prioritizes exploration, leading to unstable convergence across datasets due to frequent low-utility selections. Fixed $\epsilon=0.7$ focuses on exploitation, fails to accelerate global convergence, as it underutilizes a significant portion of experts that might have contributed to model improvement.

\begin{figure}[t]
    \centering
    \includegraphics[width=\linewidth]{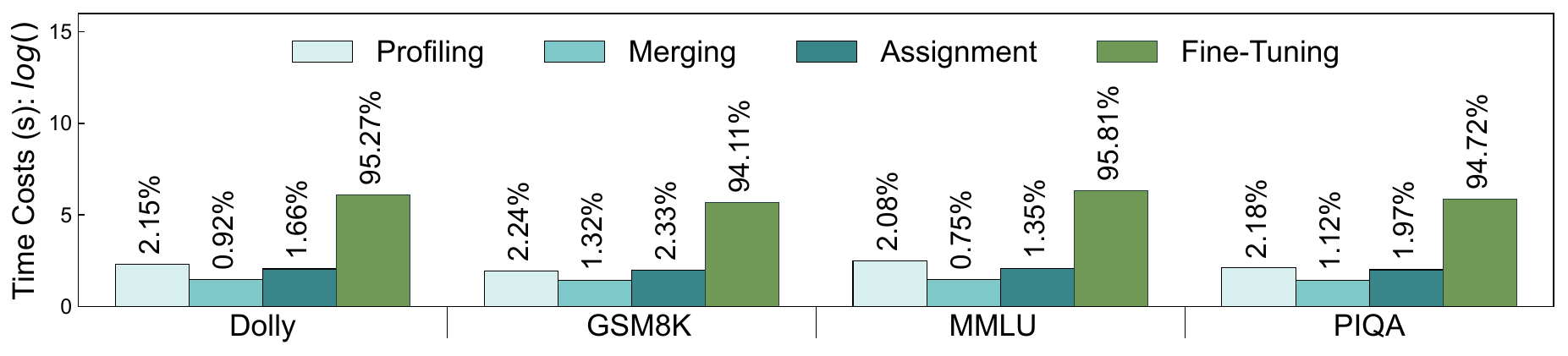}
    \caption{Additional overhead brought by \textsc{Flux}.}
    \label{fig:addition_overhead}
\end{figure}

\noindent\textbf{\textsc{Flux} overhead}.
We evaluate the computational overhead introduced by \textsc{Flux} by showing its log-scale normalized time cost in \autoref{fig:addition_overhead}.
Profiling accounts for the largest portion of the reported overhead. However, its impact is minimized as profiling runs in parallel with expert selection, reducing the overall delay. Additionally, we observe that expert merging and selection costs remain relatively stable across different datasets. For instance, on the GSM8K dataset, the time required for one round of fine-tuning is lower than on other datasets, resulting in a higher relative ratio of additional overhead. Despite these variations, \textsc{Flux} introduces only minimal computational overhead, contributing approximately 5\% of the total federated MoE fine-tuning time.

\section{Related Work}
\noindent\textbf{MoE Optimization. }Deploying MoE models is challenging due to the sheer model size~\cite{chen2025communication,kim2023mixture,yi2025edgemoe}. Existing works optimize MoE deployment through three main approaches. The first approach focuses on quantizing MoE models to lower precision, such as INT4, to reduce model size~\cite{kim2023mixture, frantar2024qmoe}. 
For example, QMoE~\cite{frantar2024qmoe} compresses the MoE model to a lower precision (e.g., 1 bit) with minimal loss in model accuracy, while QuantMoE-Bench~\cite{kim2023mixture} explores the fine-grained setting for MoE quantization. 
A second approach leverages dynamic expert offloading to serve MoE models under memory constraints~\cite{yi2025edgemoe,hwang2024pre,kong2024swapmoe,cao2025moe}. 
For instance, Pre-gated MoE~\cite{hwang2024pre} introduces a pre-gating mechanism, which aims to predict expert activation in the next layer so that the parameters can be pre-loaded accordingly. 
A third approach explores expert pruning for efficient MoE optimization~\cite{lu2024not,chowdhuryprovably,lin2024flame,xu2024mome}. 
Despite the effectiveness of these MoE deployment optimizations, they overlook the challenges of MoE fine-tuning under resource constraints. Directly applying these methods to fine-tuning can significantly degrade performance. 


\noindent\textbf{Federated MoE Learning. }Recently, MoE models have been studied in federated learning. For example, FedMix~\cite{reisser2021federated} trains an ensemble of specialized models and adaptively selects a user-specific subset of the ensemble models. pFedMoE~\cite{yi2024pfedmoe} facilitates federated learning on heterogeneous large models by assigning a shared homogeneous feature extractor and a gating network to participants. FedMoE~\cite{mei2024fedmoe} is a federated learning framework that enables participants to train heterogeneous models by selecting different experts. However, these methods either assume that the model is small enough to be fully loaded or directly exclude experts to form heterogeneous models and lead to suboptimal performance.

\section{Conclusion}
\textsc{Flux} is an efficient federated fine-tuning framework for large-scale MoE-based LLMs. \textsc{Flux} achieves adaptive MoE fine-tuning by selectively learning a subset of experts while considering resource constraints. To accelerate fine-tuning convergence, \textsc{Flux} incorporates an efficient local profiling module to profile experts for data efficiency, along with an expert role assignment algorithm to optimize MoE fine-tuning under resource limitations. In addition, an adaptive merging strategy for non-tuning experts is introduced to preserve model performance while reducing computational overhead. Extensive experiments demonstrate that \textsc{Flux} achieves superior fine-tuning speedup compared to existing approaches.

\section{Acknowledgments}
We thank all the anonymous reviewers and our shepherd, Myungjin Lee, for their insightful feedback on improving the paper. This work is supported by National Natural Science Foundation of China No. 62471383, Major Basic Research Program of Shandong Provincial Natural Science Foundation under Grant ZR2025ZD18. Peng Li is the corresponding author.

\bibliographystyle{ACM-Reference-Format}
\bibliography{ref}

\end{document}